\documentclass[runningheads]{llncs}

\usepackage{graphicx}
\usepackage{hyperref}

\usepackage{float}  
\usepackage{tabularx}
\usepackage{threeparttable}

\begin{document}

\title{Self-supervised Skull Reconstruction in Brain CT Images with Decompressive Craniectomy}

\author{Franco Matzkin\inst{1} \and 
Virginia Newcombe\inst{2} \and 
Susan Stevenson\inst{2} \and 
Aneesh Khetani\inst{2} \and 
Tom Newman\inst{2} \and 
Richard Digby\inst{2}\and 
Andrew Stevens\inst{2}\and \\
Ben Glocker\inst{3}\and 
Enzo Ferrante\inst{1}}


\authorrunning{F. Matzkin et al.}
\titlerunning{Self-supervised Skull Reconstruction in Brain CT Images with DC}

\institute{Research Institute for Signals, Systems and Computational Intelligence, sinc(i), CONICET, FICH-UNL (Argentina) \and 
Division of Anaesthesia, Department of Medicine, University of Cambridge (UK) \and 
BioMedIA, Imperial College London (UK)
}

\maketitle

\begin{abstract}
Decompressive craniectomy (DC) is a common surgical procedure consisting of the removal of a portion of the skull that is performed after incidents such as stroke, traumatic brain injury (TBI) or other events that could result in acute subdural hemorrhage and/or increasing intracranial pressure. In these cases, CT scans are obtained to diagnose and assess injuries, or guide a certain therapy and intervention.

We propose a deep learning based method to reconstruct the skull defect removed during DC performed after TBI from post-operative CT images. This reconstruction is useful in multiple scenarios, e.g. to support the creation of cranioplasty plates, accurate measurements of bone flap volume and total intracranial volume, important for studies that aim to relate later atrophy to patient outcome. We propose and compare alternative self-supervised methods where an encoder-decoder convolutional neural network (CNN) estimates the missing bone flap on post-operative CTs. The self-supervised learning strategy only requires images with complete skulls and avoids the need for annotated DC images. For evaluation, we employ real and simulated images with DC, comparing the results with other state-of-the-art approaches. The experiments show that the proposed model outperforms current manual methods, enabling reconstruction even in highly challenging cases where big skull defects have been removed during surgery.

\keywords{Skull reconstruction \and self-supervised learning \and decompressive craniectomy}

\end{abstract}

\section{Introduction}

Decompressive craniectomy (DC) is a surgical procedure performed for controlling the intracranial pressure (ICP) under some abnormal conditions which could be associated with brain lesions such as traumatic brain injury (TBI) \cite{Moon2017}. In this procedure, a portion of the skull (bone flap) is removed, alleviating the risks associated with the presence of hematomas or contusions with a significant volume of blood \cite{Galgano2017}. In order to monitor the patient's condition and potential complications from the injury, computed tomography scans (CTs) of the affected area are acquired before and after this intervention \cite{FREYSCHLAG2018}.

Previous works which study the complications that can emerge after DC suggest that the volume of the skull defect is an important parameter to evaluate the decompressive effort \cite{Tanrikulu2015,Sedney2014}. A manual method to estimate such volume was proposed by Xiao and co-workers \cite{XIAO2012205}. The authors developed a simple equation relying on three basic manual measurements which are multiplied and provide a good approximation of the real skull defect size. However, this method requires manual intervention and its accuracy is limited by the geometric approximation which does not take into account specific details of the skull shape.

Alternatively, the extracted bone flap volume could be estimated from a 3D model of the defect, which may be also useful for estimating materials and dimensions of eventual cranioplasty custom-made implants \cite{Hieu2003}. These can be used instead of the stored bone flap after DC, which has shown to carry potential complications if reused \cite{Herteleer2016}. Different methods can estimate such shapes: one strategy is to take advantage of the symmetry present in the images \cite{Huang2013}. However, it has the restriction of handling only unilateral DCs. Another simple and effective alternative could be the subtraction of the aligned pre- and post-operative CT scans, highlighting the missing part of the skull. Of course, this cannot be done if the provided data only contains post-operative images, which tends to be a common situation in real clinical scenarios.

We propose a bone flap reconstruction method which directly operates on post-operative CT scans, can handle any type of DC (not only unilateral) and is more accurate than current state-of-the-art manual methods. Our model employs encoder-decoder convolutional neural networks (CNN) and is trained following a self-supervised strategy, in the sense that it only requires images with complete skull for training, and avoids the need for annotated DC images. \\

\noindent \textbf{Contributions:} Our contributions are 3-fold: (i) to our knowledge, this is the first deep learning based model to perform skull reconstruction from brain CT images, (ii) the method outperforms the accuracy of manual and automatic state-of-the-art algorithms both in real and simulated DC and (iii) we introduce a self-supervised training procedure which enables learning skull reconstruction using only complete skull images which are more common than images with DC.

\section{Self-supervised skull reconstruction}
Our reconstruction method consists of a CNN which operates on binary skull images obtained after pre-processing the CT. We designed a virtual craniectomy (VC) procedure where full skulls are used to simulate DC patients by randomly removing bone flaps from specific areas. We used the VC to train various CNN architectures which follow alternative strategies: reconstructing only the missing flap or reconstructing the full skull and then subtracting. In the following, we describe in detail every stage of the reconstruction method.

\subsection{Pre-processing} 
\begin{figure}[t!]
    \centering
    \includegraphics[width=\linewidth]{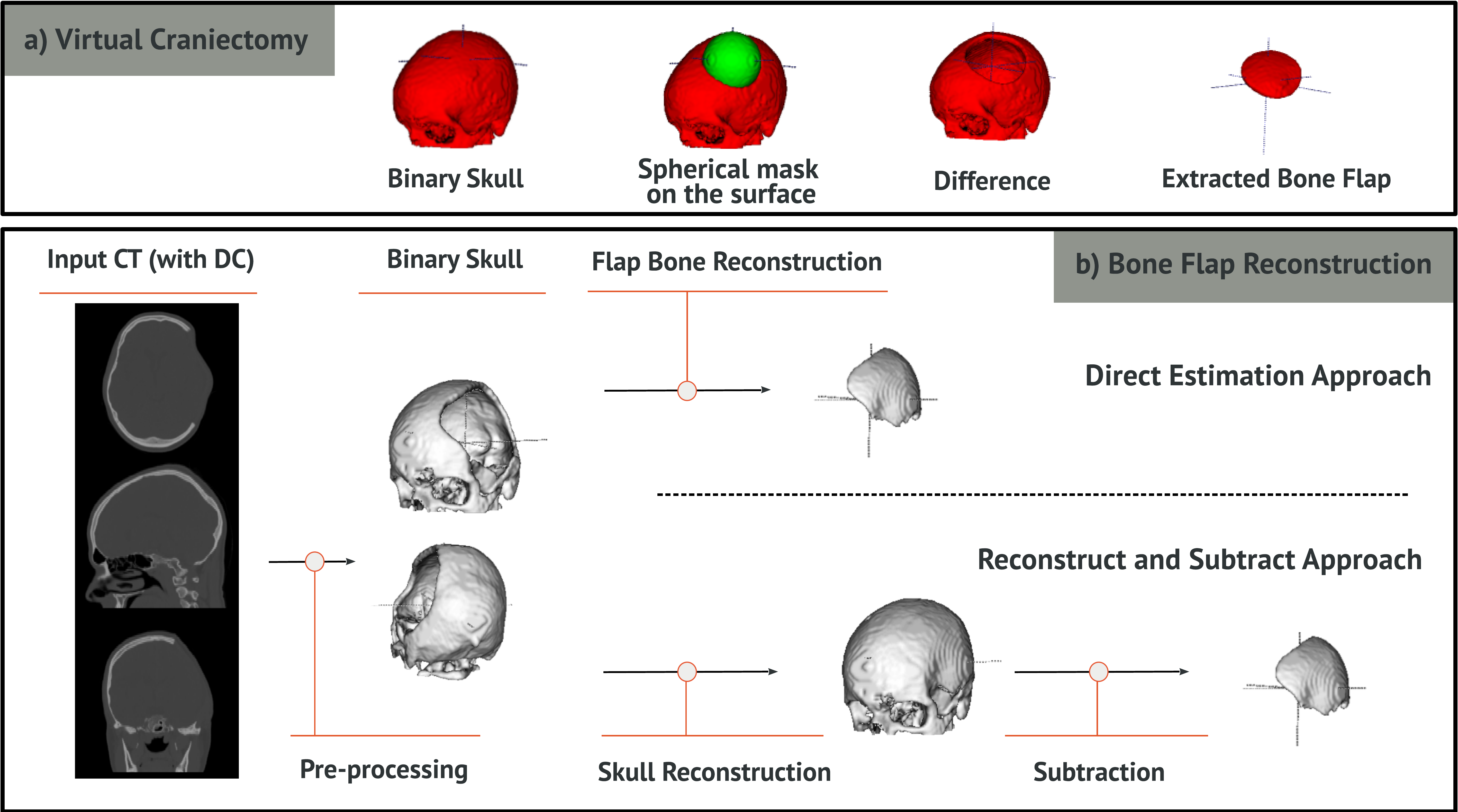}  
    \caption{a) Virtual Craniectomy process: given a skull, a spherical mask is applied in the surface for extracting a bone flap. b) In the direct estimation (DE) strategy, from the binary skull mask with DC, the bone flap is predicted by the network. In the reconstruct and subtract strategy (RS), the full skull is first reconstructed. Then, the binary mask with DC is subtracted from the complete skull, and the difference map is used as bone flap estimation.}
    \label{fig:workflow}
\end{figure}

This stage extracts a binary skull mask from a CT and consists of three steps:
\begin{enumerate}
    \item Registration: the images are registered to an atlas using rigid transformations, bringing all images into the same coordinate system. For registration, we use SimpleElastix \cite{marstal2016simpleelastix}, a state-of-the-art registration software publicly available. This pre-alignment encourages the model to focus on variations in the morphology of the skull, rather than learning features associated with its orientation and position.
    \item Resampling: After registration, images are resampled to isotropic resolution (2mm).
    \item Thresholding: In CT scans, global thresholding \cite{vanEijnatten2018}  can be employed to extract the bones due to their high values in terms of Hounsfield units \cite{seeram2015computed}. We used a threshold value of 90HU. As we can observe in Figure \ref{fig:workflow}b) a binary mask of the skull is obtained after pre-processing.

\end{enumerate}

\subsection{Virtual Craniectomy}

We designed a virtual craniectomy procedure to simulate the effect of DC on full skulls. This enables the use of head CTs with the complete skull to self-supervise the learning process, avoiding the need of manually annotated DC images where the flap is segmented. This process implies extracting the intersection of the input skull with a spherical-shaped binary mask, which can be located in its upper part and have a variable size, and use  such intersection as the ground truth during training.

We remove skull flaps from random locations, excluding the zone corresponding to the lower part (containing the bones between the jaw and the spine), where a craniectomy would not occur. The radius of the sphere was established so that the volume of the extracted bone flaps would match with standard surgeries. We defined a radius between 5 and 53 voxels to simulate craniectomies of 0.7 to 350 ${cm}^{3}$ of flap volume. This process is depicted in Figure \ref{fig:workflow}a).

\subsection{Network Architectures}
We implemented alternative encoder-decoder CNN architectures to address the flap reconstruction problem which are based on fully convolutional neural networks, but follow different reconstruction strategies, illustrated in Figure \ref{fig:workflow}b). Note that our contributions are not related to novel CNN architectures (we employ standard autoencoders and U-Net), but regarding the VC-based self-supervised strategy and its application to a new problem (i.e. skull reconstruction) where deep learning approaches have not been explored to date.\\

\noindent \textbf{\textit{a) Reconstruct and subtract with autoencoder (RS-AE):}} The first model is a fully convolutional autoencoder (AE) trained to reconstruct the complete version of a DC skull (see the Supplementary Material for a detailed description of the AE architecture). Following an approach similar to that of Larrazabal et al. \cite{larrazabal2019,larrazabal2020}, we employ a denoising AE where the training process does not only include noise for data augmentation, but also virtual craniectomies. During training, we employ only full skulls: a random VC is applied before the skull enters the AE, which is trained to output its full version. Similar to previous strategies initially developed for unsupervised lesion detection \cite{pawlowski2018}, at test time, we reconstruct the bone flap by subtracting the original DC and its reconstructed full version to generate a difference map. The difference map constitutes the final bone flap 3D estimation, from which we can compute features like volume, etc.\\

\noindent  \textbf{\textit{b) Direct estimation with U-Net (DE-UNET):}} The second model directly estimates the bone flap, avoiding the full skull reconstruction and subtraction steps, which may introduce errors in the process. We employ the same encoder-decoder architecture used for the AE, but including skip connections, resulting in a 3D version of the standard U-Net architecture \cite{unet2d} (a detailed description of the architecture is given in the Supplementary Material). For training, instead of aiming to reconstruct the full skull, we learn to reconstruct the bone flap removed during the VC. Note that, similar to the previous model, we only require full skulls for training, enabling self-supervised learning without bone flap annotations.\\

\noindent \textbf{\textit{c) Reconstruct and subtract with U-Net (RS-UNET):}} For completeness, we also explore the use of the U-Net following the reconstruct-and-subtract strategy, to evaluate the impact of the skip-connections in the resulting reconstruction. 

\begin{figure}[t]
    \centering
    \includegraphics[width=1\linewidth]{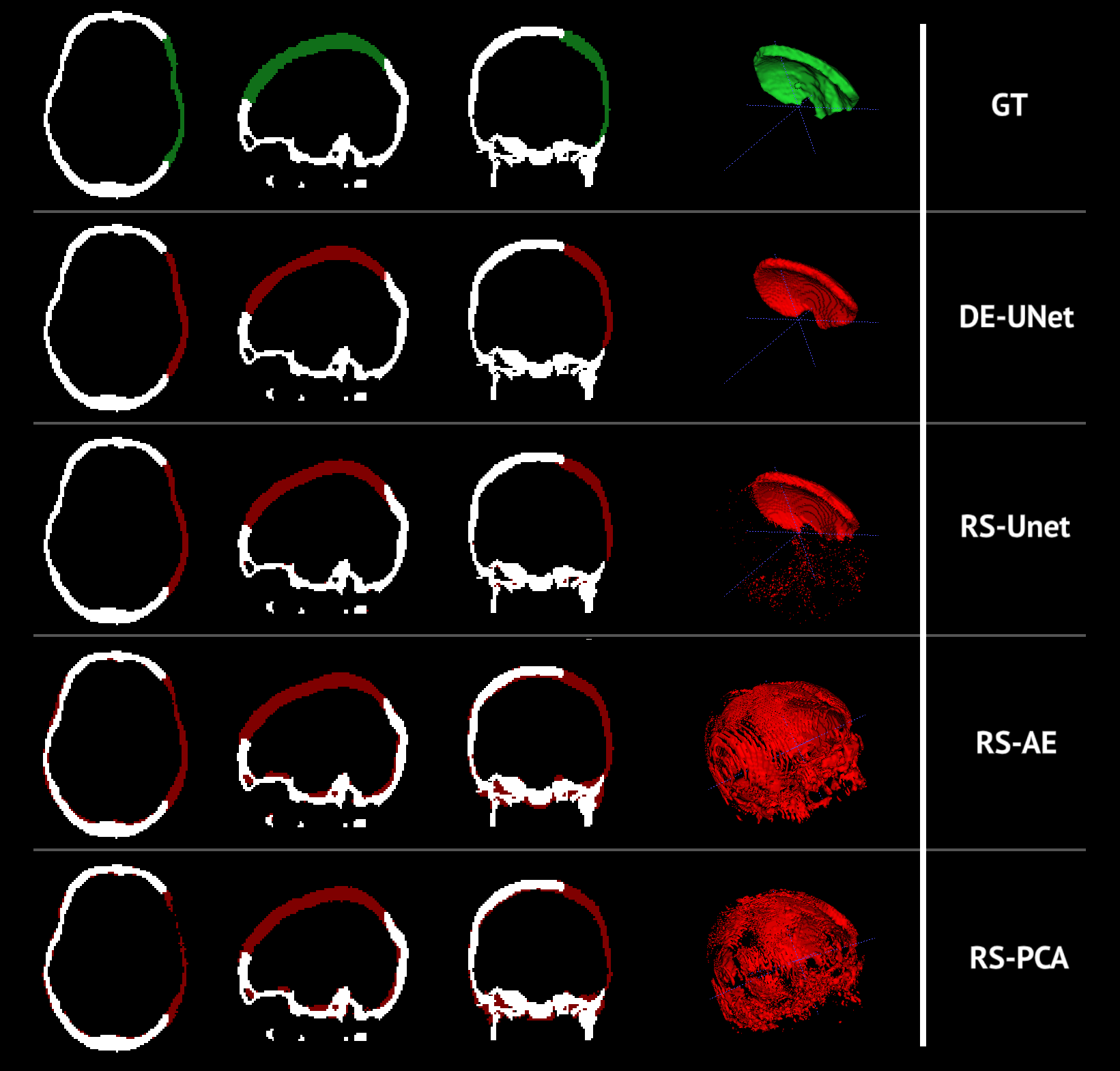}
    \caption{Bone flap reconstruction (in red) obtained with the approaches compared in this work for a real decompressive craniectomy case from our test dataset.}
    \label{fig:example-flaps-methods}
\end{figure}

\subsection{Training and Implementation}
The CNN architectures were implemented in PyTorch 1.4 and trained on an NVIDIA TITAN Xp GPU \footnote{The source code of our project is publicly available at: \url{http://gitlab.com/matzkin/deep-brain-extractor}}. During training, the images are fed to the network by adding salt and pepper noise and performing VC (with probability of 0.8), allowing the networks to see both intact and VC skulls. 
For all models the loss function $L$ consists in a combination of the Dice loss $L_{Dice}$ and the Binary Cross Entropy (BCE) Loss $L_{BCE}$. While cross-entropy loss optimizes for pixel level accuracy, the Dice loss function enhances the segmentation quality \cite{Patravali18}. The compound loss function is defined as:

\begin{equation}
    L  =  L_{Dice} + \lambda L_{BCE}
\end{equation}

\noindent where the parameter $\lambda = 1$ was chosen by grid search. To improve generalization we incorporated dropout layers and use early stopping on validation data.

\section{Experiments and Discussion}
\subsection{Database}
The images used for this work were provided by the University of Cambridge (Division of Anaesthesia, Department of Medicine). They  consist in 98 head CT images of 27 patients with Traumatic Brain Injury (TBI), including 31 images with DC and 67 cases with full skull. For training, we used full skull images only, excluding those patients who also have associated an image with DC. Patients which include pre and post-operative CT images were used for testing, since the difference between both images after registration was employed as ground-truth for the evaluation of bone flap estimation (an example is shown in green in Figure \ref{fig:example-flaps-methods}). In this context, we employed 52 images for training (corresponding to 17 different patients) and 10 for testing (since there are only 10 patients with pre and post DC studies). The 36 images not included in the study were either pre-operative images of patients from the test split or post-operative without their corresponding pre-operative.

\subsection{Baseline models}
We implemented a baseline model based on principal component analysis (PCA) for the task of flap bone estimation which follows the reconstruct and subtract strategy (RS-PCA). The principal components (see the Supplementary Material for visualizations of these components) were obtained by applying PCA to the vectorized version of the pre-processed complete skulls from the training fold. Similar to the RS-AE approach, the learnt latent representation provides a base for the space of complete skulls. Therefore, for reconstruction, we take the incomplete skull and project it to the learnt space to obtain its full version.

For the task of flap bone volume estimation, we also compared our methods with the manual state-of-the-art ABC approach \cite{XIAO2012205}. The ABC method requires to annotate manual measurements on the DC images (see the Supplementary Material for an example) and estimates the flap volume following simple geometric rules (a complete description of ABC can be found in the original publication \cite{XIAO2012205}). 

\subsection{Experiments and results}
We performed experiments for bone flap reconstruction and volume estimation in real and simulated craniectomies. The simulations were done by performing 100 random virtual craniectomes to every complete skull from the test fold, resulting in a total of 1000 simulations for test. Figure \ref{fig:example-flaps-methods} provides a qualitative comparison of the reconstructions (in red) obtained using the different approaches in a real DC. It can be observed that those based on the reconstruct and subtract strategy using AE and PCA produce spurious segmentations in areas far from the flap. The best reconstructions are achieved using the DE-UNet and RS-UNet, highlighting the importance of the skip connections. 

\begin{figure}[t!]
    \centering
    \includegraphics[width=\linewidth]{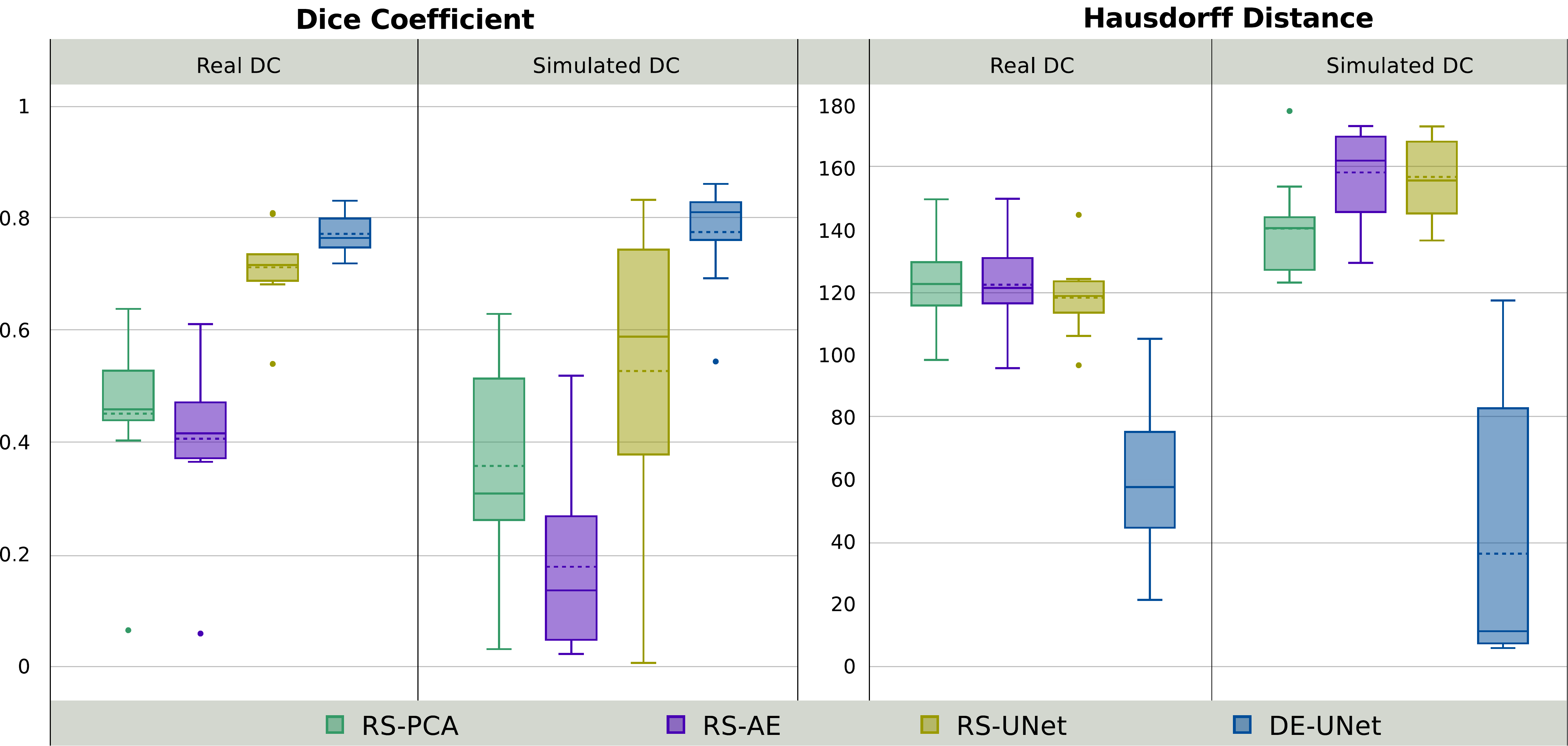}
    \caption{Dice Coefficient and Hausdorff Distance (in mm) of the proposed methods output compared with the ground-truth (dashed line indicates the mean value). It can be seen that the DE-UNet outperforms the other discussed methods.}
    \label{fig:flap-estimation-boxplots}
\end{figure}

\begin{figure}[t!]
    \centering
    \includegraphics[width=\linewidth]{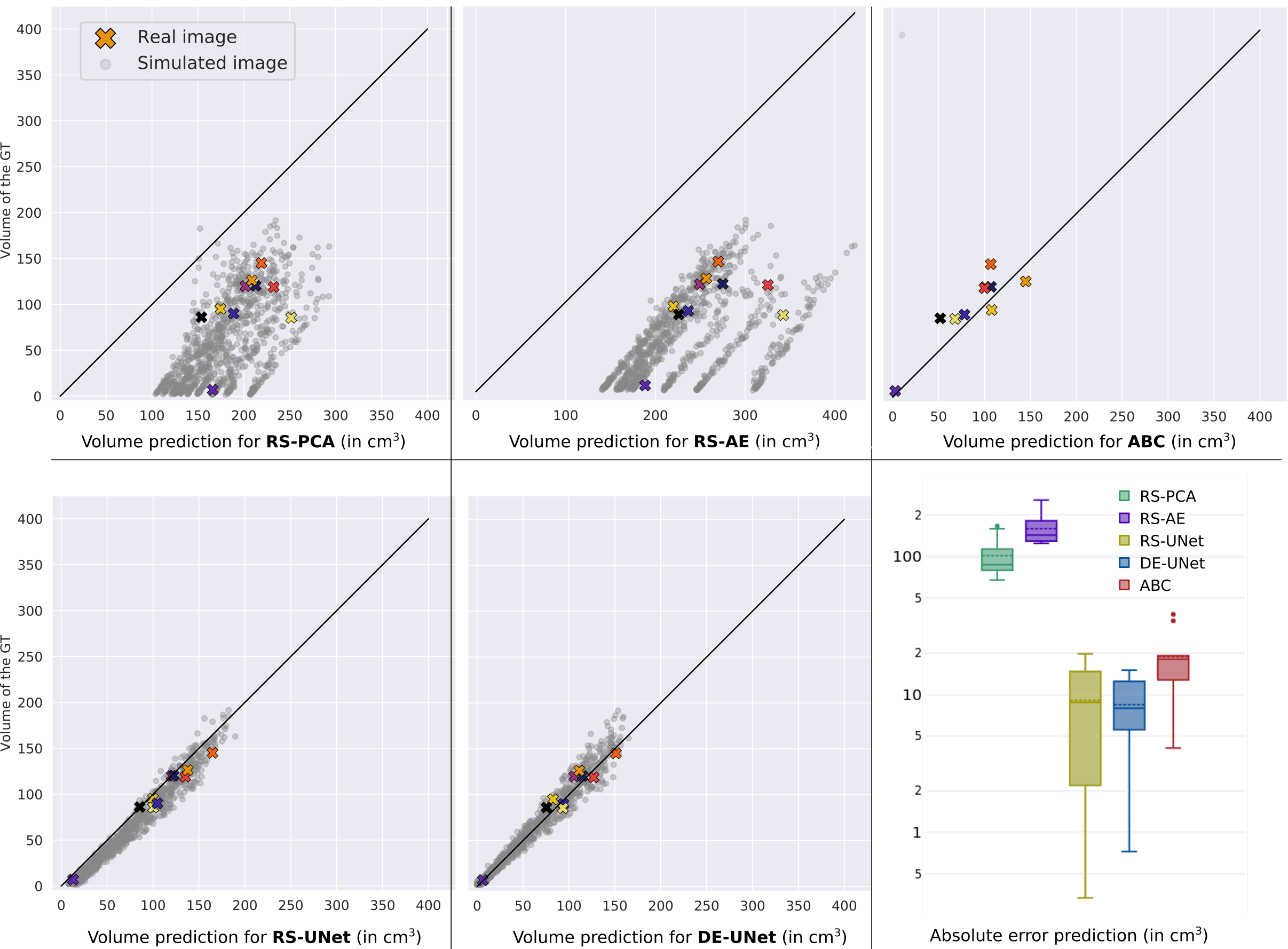}
    \caption{Quantitative comparison for bone flap volume estimation with the different methods implemented in this study. The scatter plots show the estimated (x-axis) vs ground-truth bone flap volume. Note that RS-PCA, RS-AE, RS-UNet and DE-UNet show results for both real (cross markers in color) and simulated cases (circles in grey). For ABC, we only show results in real cases since the actual CT image is required for manual annotation (and virtual craniectomies for simulations are perform directly on the binary skull mask).}
    \label{fig:volume-all}
\end{figure}

The quantitative analysis is summarized in Figures \ref{fig:flap-estimation-boxplots} and \ref{fig:volume-all}. Figure \ref{fig:flap-estimation-boxplots} shows Dice coefficient and Hausdorff distance between the ground-truth and reconstructed bone flaps for all the methods in real and simulated scenarios. Figure \ref{fig:volume-all} includes scatter plots showing the accuracy of the bone flap volume estimation: we compare the predicted volume (x-axis) with the expected volume (y-axis). The closer the points to the identity, the more accurate are the predictions. For volume estimation we also include the manual ABC method. From these results, we observe that DE-UNet outperforms the other methods in both tasks, producing even better volume estimations than the manual ABC approach. We observed that Reconstruct and Subtract methods usually generate spurious pixels as prediction (as can be seen in Figure \ref{fig:example-flaps-methods}) and a post-processing step may be needed after subtracting the pre and post-operative images (e.g. taking the biggest connected component, or applying morphological operations in the prediction). This does not tend to happen with Direct Estimation, what explains the gain in performance.

\section{Conclusions}
In this work, we propose and compare alternative self-supervised methods to estimate the missing bone flap on post-operative CTs with decompressive craniectomy. To our knowledge, this is the first study that tackles skull reconstruction and bone flap estimation using deep learning. We introduced a self-supervised training strategy which employs virtual craniectomy to generate training data from complete skulls, avoiding the need for annotated DC images. 

We studied two different reconstruction strategies: direct estimation (DE) and reconstruct and subtract (RS). We found that DE outperforms RS strategies, since the last ones tend to generate spurious segmentations in areas far from the missing bone flap. The proposed methods were also compared with a PCA-based implementation of the RS reconstruction process and a state-of-the-art method (ABC) used in the clinical practise which requires manual measurements and relies on a geometric approximation. The proposed direct estimation method based on the U-Net architecture (DE-UNet) outperforms all the other strategies.

The performance of our method was measured in real cases (TBI patients who underwent decompressive craniectomy) as well as simulated scenarios. In the future, we plan to explore the use of the bone flap features to improve patient treatment. In this sense, we are interested in studying specific features in terms of volume and shape of a craniectomy that leads to fewer complications and improves patient outcome after TBI.\\

\noindent \textbf{Acknowledgments}. The authors gratefully acknowledge NVIDIA Corporation with the donation of the Titan Xp GPU used
for this research, and the support of UNL (CAID-PIC-50220140100084LI) and
ANPCyT (PICT 2018-03907).

\bibliographystyle{splncs04}
\bibliography{mybibliography}

\newpage
\setcounter{page}{1}
\setcounter{section}{0}

\section{Supplementary material}

\subsection{Network architecture}

\begin{threeparttable}[h!]
  \begin{center}
    \caption{Architecture of the UNet model used for RS-UNet and DE-UNet.}
    \label{tab:architecture}
    \begin{tabularx}{1\textwidth}{XXXXXXXXXX}
    \textbf{Bl.} & & \multicolumn{2}{c}{\textbf{\# Kernels}} & \textbf{Act. f} &  \textbf{KS} & \textbf{St} & \textbf{BN}& \textbf{DO} & \textbf{Cat}\\
      & & \textbf{In} & \textbf{Out} &  &  &  & & &\\
      \hline
      DB1 & Co & 1 & 10 & ReLU & (5,5,5) & (1,1,1) & Yes & No&\\
          & Co & 10 & 10 & ReLU & (5,5,5) & (1,1,1) & Yes & Yes&\\
     \hline
      DB2 & Mp &  &  &  & (2,2,2) & (2,2,2) &  & &\\
          & Co & 10 & 20 & ReLU & (5,5,5) & (1,1,1) & Yes & No&\\
          & Co & 20 & 20 & ReLU & (5,5,5) & (1,1,1) & Yes & Yes&\\
     \hline
      DB3 & Mp &  &  &  & (2,2,2) & (2,2,2) &  & &\\
          & Co & 20 & 40 & ReLU & (5,5,5) & (1,1,1) & Yes & No&\\
          & Co & 40 & 40 & ReLU & (5,5,5) & (1,1,1) & Yes & Yes&\\
     \hline
      DB4 & Mp &  &  &  & (2,2,2) & (2,2,2) &  & &\\
          & Co & 40 & 80 & ReLU & (5,5,5) & (1,1,1) & Yes & No&\\
          & Co & 80 & 80 & ReLU & (5,5,5) & (1,1,1) & Yes & Yes&\\
    \hline  
      C5 & Mp &  &  &  & (2,2,2) & (2,2,2) &  & \\
         & Co & 80 & 160 & ReLU & (5,5,5) & (1,1,1) & Yes & No&\\
         & Co & 160 & 160 & ReLU & (5,5,5) & (1,1,1) & Yes & Yes&\\
     \hline
      UB6 & TrCo & 160 & 160 &  & (2,2,2) & (2,2,2) & No & No & DB4\\
          & Co & 160 & 80 & ReLU & (5,5,5) & (1,1,1) & Yes & No&\\
          & Co & 80 & 80 & ReLU & (5,5,5) & (1,1,1) & Yes & Yes&\\
    \hline
      UB7 & TrCo & 80 & 80 &  & (2,2,2) & (2,2,2) & No & No & DB3\\
          & Co & 80 & 40 & ReLU & (5,5,5) & (1,1,1) & Yes & No&\\
          & Co & 40 & 40 & ReLU & (5,5,5) & (1,1,1) & Yes & Yes&\\
     \hline
      UB8 & TrCo & 40 & 40 &  & (2,2,2) & (2,2,2) & No & No & DB2\\
          & Co & 40 & 20 & ReLU & (5,5,5) & (1,1,1) & Yes & No&\\
          & Co & 20 & 20 & ReLU & (5,5,5) & (1,1,1) & Yes & Yes&\\
      \hline
      UB9 & TrCo & 20 & 20 &  & (2,2,2) & (2,2,2) & No & No & DB1\\
          & Co & 20 & 10 & ReLU & (5,5,5) & (1,1,1) & Yes & No&\\
          & Co & 10 & 10 & ReLU & (5,5,5) & (1,1,1) & Yes & Yes&\\
          & Co & 10 & 2 & ReLU & (1,1,1) & (1,1,1) & No & No&\\
          
    \end{tabularx}
    \begin{tablenotes}
      \small
      \item \textbf{KS:} Kernel size. \textbf{St:} Stride. \textbf{BN:} Batch-Norm. \textbf{DO:} Dropout (50\%). \textbf{Cat} Concatenate. \textbf{DB:} Down-block. \textbf{CB:} Center-block. \textbf{UB:} Up-block.
      \item \textbf{Co:} 3D Convolution layer. \textbf{Mp:} 3D Max-Pooling. \textbf{TrCo:} 3D Transposed convolution layer.
      \item All models were trained for 350 epochs, using a Batch size of 9. 
      \item The AE model is equal to this one but omitting the Concatenate column.
    \end{tablenotes}
  \end{center}
\end{threeparttable}

\subsection{The ABC method}
\begin{figure}[H]
    \centering
    \includegraphics[width=.62\linewidth]{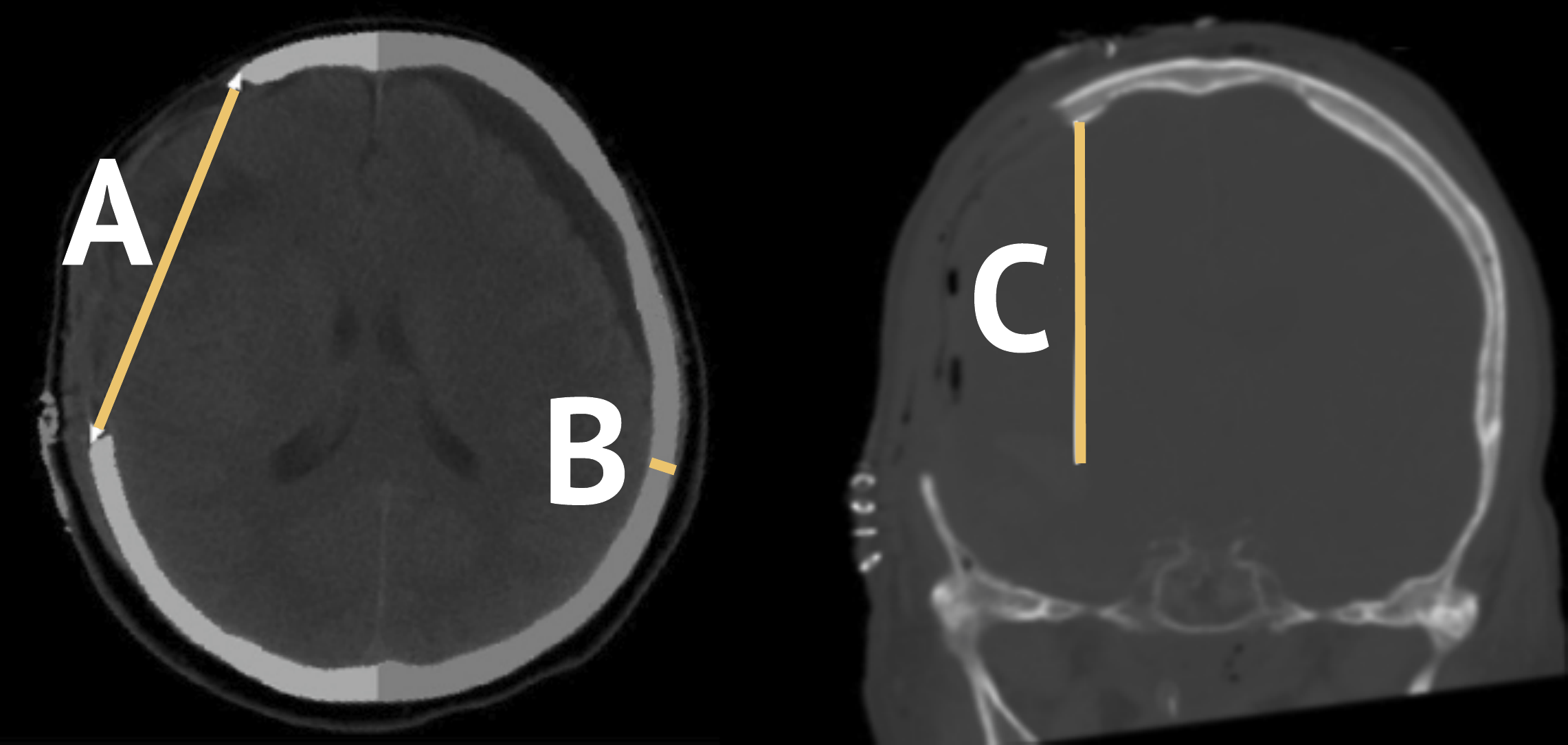}
    \caption{The manual method proposed in \cite{XIAO2012205} estimates the skull defect (SD) volume $V=ABC$, taking A as the linear distance between corners of the outer table of the SD, B as the maximum thickness measured perpendicularly to  $A$ and $C$ the sum of the inter slice distances on which full-thickness SD is visible.}
    \label{fig:abc}
\end{figure}

\subsection{Skull reconstruction with PCA}

\begin{figure}[H]
    \centering
    \includegraphics[trim={0 11.6cm 0 0},clip,width=.6\linewidth]{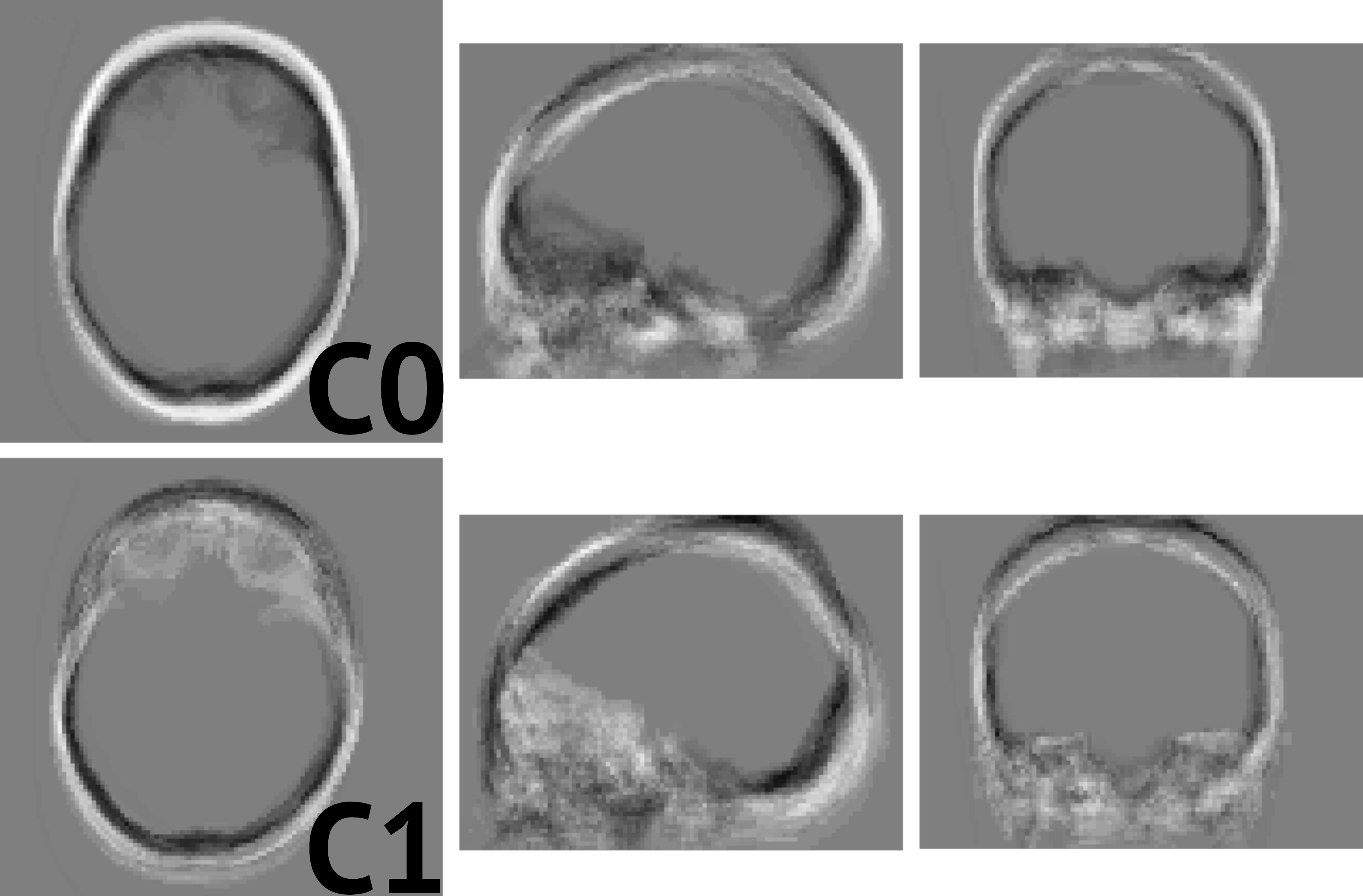}
    \caption{For this work, the PCA transformation was taken to the training split, which consists in images of patients without DC. The test images are then projected into this space and the inverse transformation is taken for going back to the image space. This image shows the visualization of the first two components.}
    \label{fig:eigenfaces}
\end{figure}

\end{document}